\documentclass[conference, 10pt]{IEEEtran}

\usepackage[cmex10]{amsmath} \interdisplaylinepenalty=2500
\usepackage{cite}
\usepackage{flushend}
\usepackage[pdftex]{graphicx}
\usepackage[bookmarks=false,hidelinks]{hyperref}
\usepackage{subcaption}
\usepackage{tikz}
\usepackage{xcolor}

\usepackage[a-1b]{pdfx}

\newcommand{\IPVAT}{TCID}
\newcommand{\IPVATGITHUB}{\url{https://github.com/Tribler/py-ipv8}}
\newcommand{\Tribler}{Tribler}

\newcommand\copyrighttext{%
\footnotesize \copyright 2021 IEEE. Personal use of this material is permitted. Permission from IEEE must be obtained for all other uses, in any current or future media, including reprinting/republishing this material for advertising or promotional purposes, creating new collective works, for resale or redistribution to servers or lists, or reuse of any copyrighted component of this work in other works. Final, published version DOI: \url{https://doi.org/10.1109/LCN52139.2021.9525011}}
\newcommand\copyrightnotice{%
\begin{tikzpicture}[remember picture,overlay]
\node[anchor=south,yshift=10pt] at (current page.south) {\fbox{\parbox{\dimexpr\textwidth-\fboxsep-\fboxrule\relax}{\copyrighttext}}};
\end{tikzpicture}%
}

\begin{document}
\begin{NoHyper}

\title{A Truly Self-Sovereign Identity System\vspace{-0.2em}}
\author{%
\IEEEauthorblockN{Quinten Stokkink, Georgy Ishmaev, Dick Epema and Johan Pouwelse}
\IEEEauthorblockA{Delft University of Technology, The Netherlands \\
Email: \{q.a.stokkink, g.ishmaev, d.h.j.epema, j.a.pouwelse\}@tudelft.nl}
}
\maketitle
\copyrightnotice

\begin{abstract}
Existing digital identity management systems fail to deliver the desirable properties of control by the users of their own identity data, credibility of disclosed identity data, and network-level anonymity.
The recently proposed Self-Sovereign Identity (SSI) approach promises to give users these properties.
However, we argue that without addressing privacy at the network level, SSI systems cannot deliver on this promise.
In this paper we present the design and analysis of our solution \IPVAT{}, created in collaboration with the Dutch government.
\IPVAT{} is a system consisting of a set of components that together satisfy seven functional requirements to guarantee the desirable system properties.
We show that the latency incurred by network-level anonymization in \IPVAT{} is significantly larger than that of identity data disclosure protocols but is still low enough for practical situations.
We conclude that current research on SSI is too narrowly focused on these data disclosure protocols.
\end{abstract}

\begin{IEEEkeywords}
self-sovereign identity, identity management systems, peer-to-peer, privacy, pseudonymity, anonymity
\end{IEEEkeywords}

\maketitle

\section{Introduction}

In its full generality, the problem of digital identity management is one of the most important and hardest problems associated with large-scale digital infrastructures.
On the one hand, digital identities are critical elements in the pervasive digital infrastructures we use in daily life for economic activity, access to healthcare and public services, etc., but on the other hand,
data breaches, data leaks and privacy violations are all too common with the current generation of digital identity management systems~\cite{dhamija_seven_2008}.
These issues can at least partially be attributed to the reliance of these systems on centralized trusted third parties to manage all aspects of users' identities.
Over the last few years, the concept of decentralized management of digital identities labelled \emph{Self-Sovereign Identity} (SSI) has emerged that promises users control over how to collect, store and share their own identity data~\cite{wang2020self}.
This paper presents the design, implementation and evaluation of a complete, open-source, truly Self-Sovereign Identity System.

SSI systems minimize trust in third parties and assume a decentralized infrastructure for private storage by the identity holders or \emph{subjects} of credentials,
which are confirmed pieces of identity data.
Trusted third parties act only as \emph{issuers} of credentials on request by subjects and cannot learn with whom or when subjects share their credentials.
Third parties cannot learn anything about subjects except for the credentials that are explicitly shared with them.
The SSI approach, in theory, has the strong privacy guarantees of \emph{data minimization} and \emph{private data control}~\cite{allen2016path} as subjects can use any data disclosure protocol they like for selectively sharing credentials and for implementing any cryptographic or data obfuscation algorithms on the corresponding data. 

The SSI approach is unique in its ability to serve as a digital analog for identification in the physical world.
What constitutes the strength of identification in the physical world is that it (a) is always presented by its owner (e.g., showing a passport), (b) is legally and practically recognized as a valid proof of identity (e.g., by verifying the passport photo) and (c) is only shared between the identity owner and the verifier without knowledge of any other party (e.g., the state that issued a passport does not know whom it is presented to). 
The digital analog of these desirable physical properties are the digital system properties of (a) \emph{Self-Sovereignty}, (b) \emph{Credibility}, and (c) \emph{Network-level Anonymity}.         

Though promising as a concept, current research on SSI is limited in its focus on data disclosure protocols for maintaining the Self-Sovereignty of data.
However, these protocols are definitely not sufficient on their own to deliver strong privacy guarantees in practical applications.
This is evident from the research on privacy-focused decentralized systems such as anonymous crypto-currencies~\cite{fanti_dandelion_2018}, which shows that not only protocol-level privacy, achievable with cryptographic tools, but also network-level anonymity is necessary for practical privacy-preserving applications~\cite{biryukov2019deanonymization}.
Network-level anonymity, that is, obfuscating the source-destination pairs of messages for credential creation and sharing, is crucial in SSI systems since Internet traffic correlation can completely undermine the effect of the data-disclosure protocols~\cite{stopar2019emmy}.

The design of our SSI system is based on 7 functional requirements that we deduce from the three system properties it needs to provide,
which relate to basic trustworthy messaging in a communication substrate and to network-level anonymity for credential creation and verification
in two peer-based overlays on top of this substrate.
Our design is modular in that it allows for the specialization of the system for applications with differing
privacy and security requirements by incorporating different software modules for several functionalities. 
In particular, we discuss the specialization of our design for use as a digital analog of a passport. 

We are not the first to propose using network-level anonymity for Self-Sovereign Identity systems~\cite{stopar2019emmy}, but we are the first to create a viable prototype in a truly Self-Sovereign, zero-server manner.
The contributions of this paper are:
\begin{enumerate}
\item We provide a complete and substantiated modular system design for the creation of Self-Sovereign Identity solutions with strong privacy called TrustChain IDentity, or \IPVAT{} for short  (Section~\ref{sec:design}).
\item We present the specialization of \IPVAT{} for a passport-grade Self-Sovereign Identity solution (Section~\ref{sec:implementation}).
\item We perform a performance analysis of \IPVAT{} that shows that the overhead of \IPVAT{} is not prohibitive for practical use (Section~\ref{sec:evaluation}).
\end{enumerate}

\section{Problem statement}
\label{sec:problemstatement}

\begin{figure}
\centering
\includegraphics[width=\linewidth]{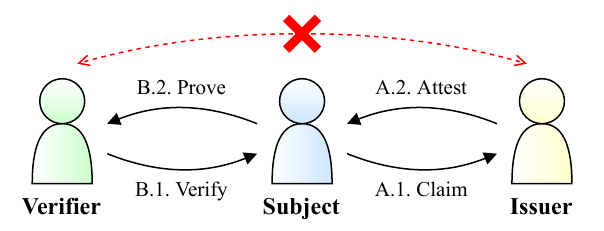} 
\caption{The two communication flows, A and B, when enrolling and verifying a credential: both flows involve the subject, as communication between the verifier and issuer violates the Self-Sovereignty property.}
\label{fig:ssiinteractions}
\end{figure}

The main research problem addressed in this paper is ``What components should the architecture of a Self-Sovereign Identity Management System leverage to provide the properties of \emph{Self-Sovereignty}, \emph{Credibility}, and \emph{Network-level Anonymity}?''
Secondarily, the architectural components required for guaranteeing Credibility and Network-level Anonymity may have an impact on the latency of attesting to and verifying credentials, which would seriously hamper the practical use of an SSI system.
Therefore, the second research question is whether the overhead incurred by these components is prohibitive for practical use or not.

The \emph{Self-Sovereignty} property, which is defined as identity data control and private data minimization by the subject, means that subjects can build their own sets of multiple pieces of identity data (their \textit{credentials}) by having issuers attest to them, and can selectively prove credentials to verifiers when performing transactions in a digital infrastructure.
For this purpose, an SSI solution should only allow the following two types of communication flows between issuers, subjects and verifiers, with four types of messages (see \autoref{fig:ssiinteractions}):
\begin{itemize}
\item \textbf{Flow A. Enrolling a credential:} a request for credential creation by a subject (A.1) and the attestation for this credential by an issuer (A.2).
\item \textbf{Flow B. Verifying a credential:} a request for proof of a credential by a verifier (B.1) and the subsequent proof of this credential by the subject (B.2).
\end{itemize}
\noindent As indicated in \autoref{fig:ssiinteractions}, a direct data flow between the issuer and the verifier is not permitted. 
An analogous physical example of these digital interactions is a citizen claiming to be older than 21 (A.1) and a government issuing a passport that proves this claim (A.2).
The citizen in this example could then later attempt to buy liquor in a liquor store which would ask for proof that this citizen is over 21 (B.1) for which the citizen shows his passport (B.2).
The government (the issuer) does not communicate with the liquor store (the verifier).

The second property of \emph{Credibility} is defined as credentials being legally and practically recognized as valid proofs of identity data.
The two cornerstones of credibility assurance in the digital domain are \emph{digital signatures} and \emph{user authentication}, which we now discuss in turn.

Digital signatures by legitimate authorities, i.e., digital signatures by third parties that are trusted by both a subject and a verifier, serve to attest to the credibility of presented claims (A.2 in~\autoref{fig:ssiinteractions}). 
Acting as an issuer, such an authority signs a claim to provide a cryptographically verifiable proof of credibility, forming a presentable credential.
For example, subjects may present a credential that shows their age and is digitally signed by a government, an issuer that most subjects and verifiers trust to be a credible source of age information.
Subjects have the ability to collect signatures of multiple authorities to further enhance the credibility of a credential, similar to how users sign each other's keys to establish credible identity in the PGP web of trust \cite{zimmermann1995official}.
However, this transitive credibility is not a one-way relationship from authorities to credentials.
Instead, the credibility of a credential also influences the legitimacy---and thereby the credibility---of authorities~\cite{grolin1998corporate}.
Credentials with provably incorrect data result in a damaged reputation of the authority who signed them.
Thus, issuers must be able to either retroactively change the data they signed or be able to revoke signatures they granted to subjects to uphold their own credibility.

The second cornerstone of credibility is user authentication of the subject.
While an interaction with a subject may be authentic and covert, device theft may have occurred: the holder of the device might not correspond to the enrolled identity.
This means that in addition to deriving credibility from the signing authority, credibility is derived from the authentication of the user holding the device, which allows an application to bind the physical user to a device.

The third and final property of SSI management systems is \emph{Network-Level Anonymity}.
Existing work shows that an adversary who is able to observe traffic in a network 
can detect a pattern of a subject's associations that may lead to identity correlation~\cite{dyer_peekaboo_2012}.
In the context of SSI systems, without network-level anonymity, the credential enrollment and verification traffic of subjects can be observed, which,  even though the data in the messages is encrypted, makes them 
vulnerable to identity correlation attacks\footnote{Though bad for users' privacy, a single centralized attesting authority avoids this, as no social network can be derived (i.e., all users will only ever interact with a single party that signs their data, the centralized authority, making all interaction graphs equal).}~\cite{srivatsa2012deanonymizing}.
Thus, an SSI management system needs to provide anonymity guarantees at the network level to preserve the protocol-level privacy guarantees.

\section{System Design}
\label{sec:design}

In this section we present the system components that jointly deliver the \emph{Self-Sovereignty}, \emph{Credibility}, and \emph{Network-level Anonymity} system properties as implemented in our open-source system \IPVAT{}, which provides a framework that enables SSI-based messaging between applications that require identification.
These applications may range from using a mobile phone to buy alcoholic beverages in a bar to a pre-flight registration over the Internet.
Without loss of generality, for all applications we treat the subject, issuer and verifier as separate entities using our system, though in practice the issuer and verifier are often the same entity (e.g., bank account credentials being used to authenticate with the same bank).
To guide the introduction of \IPVAT{}, in \autoref{fig:stack} we show its system components, distributed across a communication substrate and peer-to-peer-based overlays. 
These components jointly satisfy the \emph{7 functional requirements} detailed below that we identified for enabling the SSI interactions for credential enrollment and verification over the Internet.
The communication substrate can be implemented on top of different communication media, and in addition to our Python-based Internet implementation, we have also created a Kotlin-based Bluetooth version.

We believe \IPVAT{} to be a valid reference architecture as it can even be made to support the use case of digital passports, which has arguably the most challenging set of design requirements.
\IPVAT{} has been created in tight collaboration with both government and industry.
This was necessary to satisfy anticipated future legislation on the storage and use of identity data in Self-Sovereign Identity solutions.
However, authentication levels for digital identities have long since been standardized by the National Institute of Standards and Technology~\cite{nist2017idguidelines}.
\IPVAT{} can be specialized for specific applications by enabling the corresponding authentication levels in a modular and transparent way, as discussed for \emph{passport-grade} identities in \autoref{sec:implementation}.

\begin{figure}
\centering
\includegraphics[width=\linewidth]{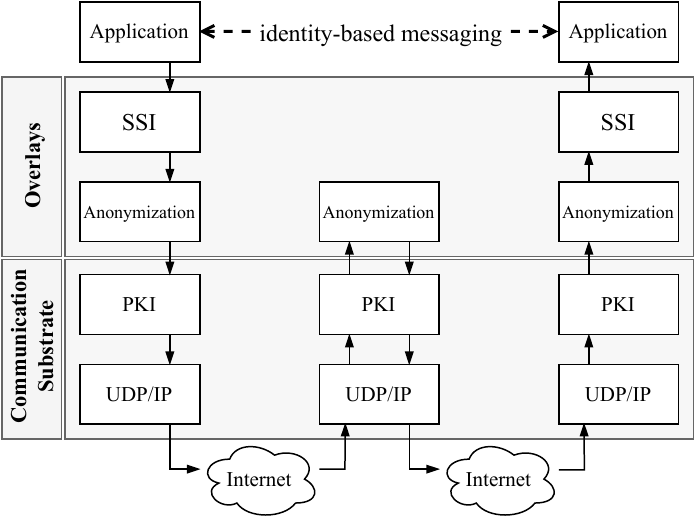} 
\caption{The data flow through the \IPVAT{} components for applications with one intermediary for anonymization, using UDP/IP over the Internet for the communication substrate.}
\label{fig:stack}
\end{figure}

Regardless of its specialization, the communication substrate of \IPVAT{} (shown in \autoref{fig:stack}) serves three functional requirements for messaging.
First, peers must be able to \emph{(1) communicate} with other peers directly, i.e., without any specific trusted third party, which can be solved using the address space of a medium (like the IPv4 and IPv6 protocols).
This direct communication circumvents the need for centralized portals that manage user identities.
Secondly, it must be impossible for peers to impersonate each other: messages should be \emph{(2) authentic}.
Lastly, the communication substrate should ensure messages are not modified by anyone other than their originator: messages should maintain \emph{(3) integrity}.
The second and third requirements are implemented using digital signatures from decentralized PKI (public key infrastructure). 

Having a communication substrate for end-to-end messaging is not enough for all possible uses of digital identity and \IPVAT{} provides two network overlays to allow further strengthening of its identification mechanism.
If attackers are able to identify users based on their messaging patterns, the security assumptions of SSI solutions are violated~\cite{stopar2019emmy}.
For an attacker can then both disrupt communication and track users through their messages. 
Practically, these attacks may be as simple as sending a large number of messages or tracking IP addresses.
Therefore, the first overlay on top of the communication substrate in \autoref{fig:stack} facilitates decentral \emph{(4) network-level anonymization}.

The anonymization overlay uses other peers in the overlay network to create covert multi-hop communication channels (\autoref{fig:stack} shows a two-hop channel).
Other peers are randomly selected, e.g., like routers in the Internet, and are therefore neither specific nor trusted third parties.
If no peers are available as intermediaries, the anonymization is essentially reduced to an expensive form of link encryption and requires the communication medium itself to not leak device identifiers.
For creating the multi-hop channels, \IPVAT{} implements its own derivative of the Tor protocol~\cite{dingledine2004tor}.
This customization is necessary as anonymization---while aiding in averting attacks~\cite{stopar2019emmy}---cannot be applied as an afterthought to existing protocols~\cite{biryukov2015bitcoin}.
Tor's ``hidden services'' protocol normally requires centralized \emph{directory servers} to share the peers that can be used for constructing channels~\cite{overlier2007improving}.
As these directory servers are again specific and trusted third parties, which \IPVAT{} wants to avoid, we have replaced them with a gossiping protocol over the anonymization overlay.
Peers in \IPVAT{} are able to \emph{(5) decentrally synchronize} the information normally published by directory servers.
However, this gossiping approach is classically vulnerable to both the Index Poisoning attack and the Eclipse attack through fake identities called Sybils.
\IPVAT{} uses a method of testing physical resources (using latency) to avoid these Sybil attacks in its anonymization overlay~\cite{wang2005efficient}.

Peers use the SSI overlay of \IPVAT{} to disclose their credentials over covert channels, i.e., to \emph{(6) identify} themselves in a Self-Sovereign fashion.
For the most stringent use cases of SSI, the overlay also offers functionality for \emph{(7) accountability} of subjects, e.g., for a government to identify an individual in case of overstaying a visa after crossing a border.
The SSI component facilitates storage for subjects and verifiers of the data structures that allow for identification and accountability: \emph{public key pairs}, \emph{credentials} and \emph{pseudonyms} (explained in \autoref{sec:implementation}).
Furthermore, the SSI overlay handles the SSI-related data flows of \autoref{fig:ssiinteractions} between peers.
The identification and accountability functionalities are exposed to the application layer, providing identity-based messaging.

\section{Passport Grade Specialization of \IPVAT{}}
\label{sec:implementation}

In this section we further specify how the SSI component of \IPVAT{} can be configured and used to support various levels of authentication, up to the point of being able to support use cases that normally require a physical passport.
This section discusses the immutable data structures that we call pseudonyms, how to check if a user is physically present when the pseudonym is used, how audit logs can be formed in a privacy-preserving manner for legal compliance and how authorities can revoke attestations previously given to users.

\subsection{Identification and Pseudonyms}
\label{sec:pseudonymsandidentification}

\begin{figure}
\centering
\includegraphics[width=\linewidth]{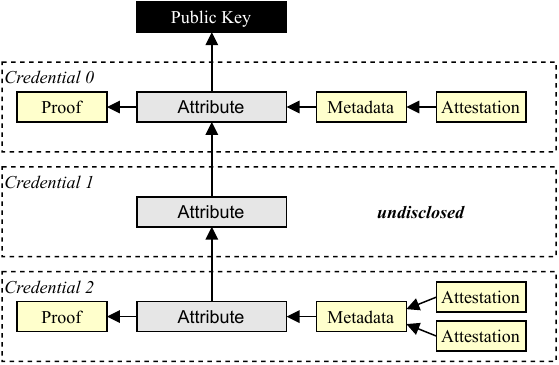} 
\caption{A pseudonym data structure corresponding to the public key of a subject with two disclosed credentials.
Arrows denote an element storing the hash of the element pointed to.
Undisclosed credentials include only an attribute, whereas their disclosure also includes a proof, metadata and attestation(s).}
\label{fig:pseudonymstructure}
\end{figure}

In \IPVAT{}, identities are implemented as pseudonym data structures, which are stored on the devices of the corresponding subjects.
As \IPVAT{} is Self-Sovereign, a pseudonym can be created by generating a key pair with any of the supported elliptic curves, without permission of a third party.
A pseudonym can also be invalidated or removed autonomously by a subject, though its use may leave a permanent record (\autoref{sec:accountability}).
The pseudonym data structure itself holds a list of credentials that are hidden from verifiers until a user wishes to disclose them.
We show an example of a pseudonym presented by a subject to a verifier in \autoref{fig:pseudonymstructure}, to guide the explanation of its elements.
\autoref{fig:pseudonymstructure} shows that a credential contains a \emph{proof}, an \emph{attribute}, \emph{metadata} and \emph{attestations} when disclosed, and only contains an \emph{attribute} when undisclosed.
When included, these elements make authentication progressively stronger up to serving as a digital equivalent of a physical passport.

Disclosure of credentials is rooted in \emph{proofs}.
Each credential has one proof.
Typically, a proof will consist of encrypted information (e.g., Pedersen commitments~\cite{pedersen1991non} and unlinkable signatures~\cite{bichsel2009cryptographic}).
Given an agreement between the subject and the verifier on a disclosure protocol for a particular proof (normally implicit, but captured explicitly in the metadata of each credential in \IPVAT{}), some property of the encrypted information can be shown to hold.
For example, one may prove the input data for a proof was larger than 21.

As shown in \autoref{fig:pseudonymstructure}, a subject can add credentials to their pseudonym by attaching \emph{attributes} using the hash of the preceding attribute or the public key of the pseudonym, if there is no preceding attribute.
Each of these attributes consists of only two hashes: a hash that points to the previous attribute or public key and a hash that points to a proof.
By creating an immutable linked list of attributes, we allow for attributes to depend on each other.
For example, the validity of a driver's license may critically depend on being enrolled with a valid name, but the name may not always have to be disclosed.
By extension, the chain of attributes implies a partial ordering of credentials, which ensures that credentials cannot be mixed-and-matched from multiple pseudonyms.
To continue our driver's license example: it is important that the presented valid name is not just any valid name, but the exact valid name that was used to enroll the driver's license.

Metadata will have to support a credential, at least to identify the data disclosure protocol for its contained proof.
To connect this metadata to an attribute and its proof, we once again use a hash to point to the respective attribute.
Finally, the collected attestations made by issuers point to the metadata of credentials.
Through the chain of hashes, this binds an attestation to the metadata of the attested credential, the proof of the credential, and all preceding credentials.
Notably, a user cannot change a proof without invalidating all attestations over credentials further down in the chain.
This mechanism allows verifiers to detect that the conditions for an issuer to attest to an attribute may have been violated.
For example, a name change automatically invalidates a government's attestation of a driver's license that is only valid for that particular name.

Attestations are digital signatures made by issuers and form one of the cornerstones for \emph{Credibility} (\autoref{sec:problemstatement}).
However, these digital signatures are made by public keys and form pseudonyms themselves.
We envision a future in which subjects require issuers to use this pseudonym to identify themselves before attesting.
For example, an issuer's pseudonym must first prove to work for the government before attesting to a driver's license of a subject.
However, which pseudonyms to trust when engaging in business logic is up to users.
For the standardization of business logic surrounding pseudonyms with attributes, we refer to the NIST recommendations on Attribute-Based Access Control~\cite{hu2014guide}.

\subsection{Proof of user presence}

Just like attestation, user authentication is one of the two cornerstones of \emph{Credibility} in SSI systems (\autoref{sec:problemstatement}).
Establishing a communciation session between peers using just public keys is not enough for a passport use case.
The user who created a pseudonym should prove they are present when using it, otherwise device theft could lead to identity theft.
To this end, a communication session established by the communication substrate can be further strengthened by using credentials for authentication~\cite{kurmi2015survey}.
Such an authenticating credential functions similarly to proving ownership of a public key with a digital signature: a subject shows a verifier a publicly verifiable property of private data they hold.

The contents of authenticating credentials can vary from passwords to biometric checks, and Physically Unclonable Functions (PUFs)~\cite{niya2020kyot}.
The more authentication credentials are added to a pseudonym, the better its security guarantees will be (potentially even beyond the standards of the NIST~\cite{nist2017idguidelines}).
The first deployed prototype of \IPVAT{} included only closed-source facial recognition with liveness detection.

\subsection{Accountability and Audit Logs}
\label{sec:accountability}

In certain use cases, like border crossing, verifiers are legally required to keep audit logs.
Such logs exist for governments to hold individuals \emph{accountable} for their use and misuse of their identities~\cite{accorsi2008automated, peyton2007audit}.
Audit logs can be formed by the verifier storing the disclosed credentials presented by users~\cite{deploy2018stokkink}.
The digital signatures and hashes that constitute a pseudonym make the audit logs immutable, to avoid tampering and to hold legal status.
However, logs are not shared publicly.
Only certified auditors with the appropriate legal grounds may require verifiers to share an audit log.
For example, immigration services may audit border authorities.

To uniquely identify users, \IPVAT{} offers privacy-preserving credential construction that allows legally required checks by certified auditors.
This feature is modular and can be enabled to provide the full range of passport grade use cases.
Before engaging in verification of a credential, subjects present verifiers with a credential that is attested to by a certified auditor without proving the credential.
This special credential contains an encrypted reference to the natural person that is captured in a government database.
Through the auditor maintaining a cryptographic secret link between the attribute of a pseudonym and a central register of all citizens, a pseudonym can be traced back to the enrolled individual by the auditor.
The implementation of this optional credential is transparent and limited to specific use cases grounded in appropriate legal frameworks.
In the spirit of Self-Sovereignty, both the subject and the verifier may opt not to include or use the credential used for auditing, though it may not be legal to do so.

\subsection{Revocation by Authorities}
Several solutions have been proposed to enable revocation by authorities of the subject credentials they have signed~\cite{hajny2012unlinkable}.
We discuss three of them.
The first and fastest solution adds a link to the revocation register (central server) of the issuer, e.g., a ``revocation authority'' in IRMA~\cite{lueks2017fast}.
However, this issuer will have to be online to allow access to the revocation register and forms a specific and trusted third party---which is not Self-Sovereign.
The second solution is the complete decentralization of revocations using a shared log, e.g., a blockchain---as used by Sovrin~\cite{khovratovich2017sovrin}, using a cryptographic construction to check for revocation in a privacy-preserving manner.
However, writing to a shared log takes more time for the revocation to reach finality, can invalidate the incentives the log was built around and has the issue of unbounded growth of the log.
Finally, the third solution is to include validity terms into the metadata of credentials, e.g., ``epochs of lifetime'' in Idemix~\cite{bichsel2009cryptographic}.
As a credential is then only valid for a limited amount of time, the chance of a credential being revoked is related to the duration of its validity.
This approach requires frequent re-enrollment of credentials.
None of these three available revocation methods fits all use cases, so \IPVAT{} enables all three to allow adaptation to any use case.

\section{Evaluation}
\label{sec:evaluation}

The key system property for the usability of SSI solutions in practical situations is low interaction latency.
It would be inconvenient to wait an hour at the airport to have your identity verified and, in fact, electronic border control should take no longer than 30 seconds to be considered usable~\cite{labati2016biometric}.
In this section we show that the latency of the interactions within the SSI overlay is well within this 30-second limit, finishing consistently within three seconds for all of the implementations of credential proofs we evaluate.
In contrast to the three-second latency of these SSI-overlay interactions, 
the latency of sending an SSI-overlay message across a covert channel in the anonymization overlay is shown to be five seconds and up to over 20 seconds.
We also attempt to find a metric to serve as a tie-breaker for the different implementations of the SSI-overlay interactions.
However, we have not been able to find it, as the latency, CPU usage and network traffic of the implementations are all similar.

\subsection{Experimental Setup}
We measure the performance of the enrollment and verification of credentials.
We compare the credential proof implementations of \IPVAT{} (see \autoref{sec:pseudonymsandidentification}) to those available in Hyperledger Indy~\cite{saraf2018blockchain}, uPort and Jolocom~\cite{gruner2019integration}.
These three external solutions have been identified as usable in a study by Bartolomeu~\cite{bartolomeu2019self}.
The performance metrics we measure are latency, CPU usage and network traffic.

The four credential proof implementations of \IPVAT{} that we measure use three distinct protocols.
Two implementations use a Zero-Knowledge Proof protocol that allows proofs over input data with a length of $1024$ bits and $4096$ bits, which we refer to as the ``TCID-1024'' and the ``TCID-4096'' proofs, respectively.
One implementation uses a Non-Interactive Zero-Knowledge Range Proof protocol, which we refer to as the ``TCID-PB'' proof~\cite{peng2010efficient}, and which allows subjects to prove input data lies in a certain number range.
Lasty, we measure \IPVAT{}'s implementation of the IRMA protocol~\cite{alpar2017irma} (a blinded signature scheme based on Idemix~\cite{bichsel2009cryptographic}), denoted as ``TCID-IRMA''.

We have measured the Python implementation of \IPVAT{}.
All experiments are performed on a virtual machine running Ubuntu 19.10, with 4 CPU cores fixed to 3.50 GHz and 16\,384 MB of memory.
Each of the presented boxplots is constructed from twenty data points.
Credential enrollment and verification is measured as-is with default settings, no modifications have been made to the publicly available source code.

\subsection{Latency}
\label{sec:latencybenchmark}

\begin{figure}
\centering
\includegraphics[width=\columnwidth]{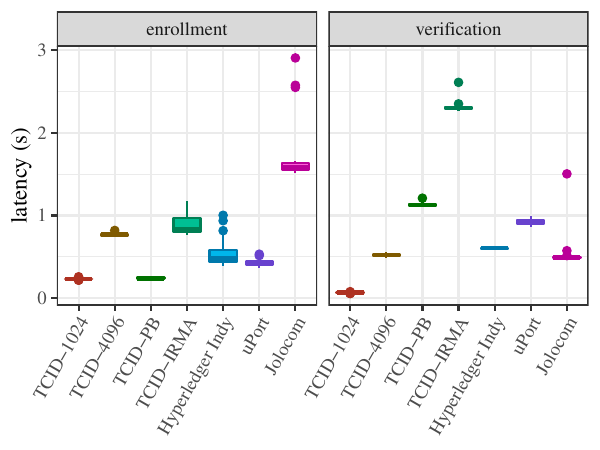}
\caption{Boxplots of the credential enrollment and verification latency for different proof implementations.}
\label{fig:perflatency}
\end{figure}

In \autoref{fig:perflatency} we present our measurements of the latency of credential enrollment and verification.
These latencies are defined as the time between the subject initiating Flow A of \autoref{fig:ssiinteractions} and receiving the corresponding attestation, 
and the time between the verifier initiating Flow B of \autoref{fig:ssiinteractions} and completing the proof, 
both without network transfer time.
From our latency comparison, we find no clear winner from the evaluated credential implementations. 
What is gained in enrollment latency is lost in verification latency and vice-versa, with Hyperledger Indy being the most consistent between these two categories (but also not the fastest).
The evaluated solutions all finish within 3 seconds, making them usable for, e.g., electronic border control, which should be finished in 30 seconds~\cite{labati2016biometric}.

The results of \autoref{fig:perflatency} show that the TCID-1024 proof provides the lowest latency for both enrollment and verification.
However, it would be unfair to claim that the TCID-1024 proof is the best solution, as it can only handle up to 128 bytes of information, while the other credential solutions can handle arbitrary-size inputs.
However, this does show that choosing the correct disclosure protocol is important to satisfy the need for either low enrollment or low verification latency.

\subsection{CPU Usage}
\label{sec:cpu}

\begin{figure}
\centering
\includegraphics[width=\columnwidth]{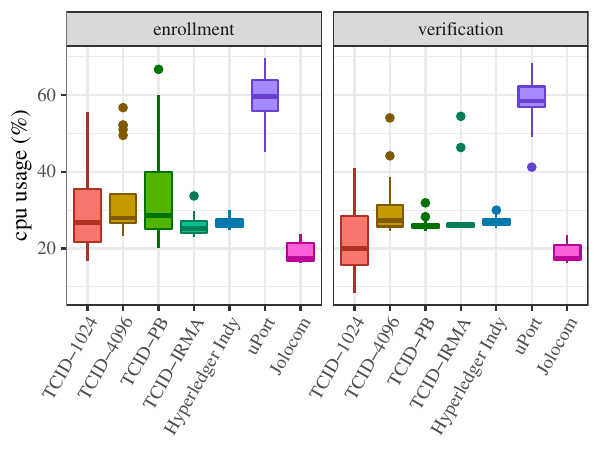}
\caption{Boxplots of the credential enrollment and verification cpu usage for different proof implementations.}
\label{fig:perfcpu}
\end{figure}

We explore CPU usage as a potential selling-point of disclosure protocols.
We measure the CPU usage as the average CPU utilization of the subject and issuer during Flow A of \autoref{fig:ssiinteractions} and of the subject and the verifier during Flow B of \autoref{fig:ssiinteractions}.
As our experiments run on a homogeneous virtualized setup, they serve as a rough estimate for the power consumption of the different Self-Sovereign Identity solutions.
As in the previous latency experiment, we measure each credential creation and verification implementation separately.
The results are visualized in \autoref{fig:perfcpu}.
Only uPort offers a surprise, in that it has a higher CPU usage for both enrollment and verification of credentials, seemingly due to its web server.
From our experiments, we infer that CPU is not a significant distinguishing factor for current Self-Sovereign Identities.

\subsection{Network Traffic}
\label{sec:nettraffic}

\begin{figure}
\centering
\includegraphics[width=\columnwidth]{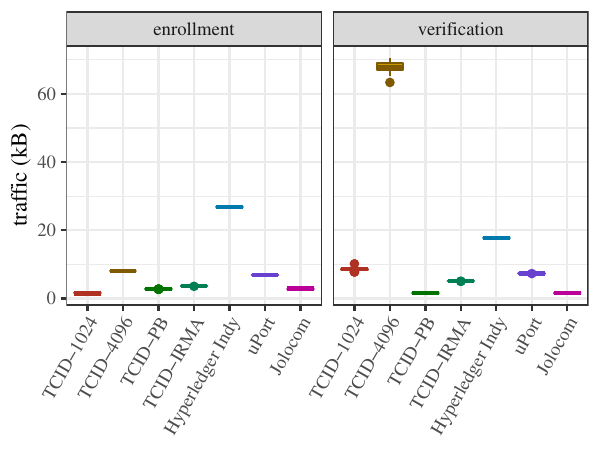}
\caption{Boxplots of the credential enrollment and verification network usage for different proof implementations.}
\label{fig:perftraffic}
\end{figure}
 
We now evaluate the network traffic as a result of the various credential implementations.
To contextualize these results, we note that, since the deployment of 3G mobile communication networks, mobile phones are able to transfer megabytes per second~\cite{ezhilarasan2017review}.
From the results of our experiment, visualized in \autoref{fig:perftraffic}, we derive that the transfer rate of several megabytes already amounts to overprovisioning for these implementations.

A notable result is that interactive Zero-Knowledge Proofs generate much more traffic than Non-Interactive Zero-Knowledge Proofs and signature schemes to verify.
The worst case, TCID-4096 proof, requires up to 70 kilobytes of data to be transferred (though this is still completely acceptable for devices capable of communicating at several megabytes per second).
The low traffic for Non-Interactive Zero-Knowledge Proofs is expected, as it is their design goal~\cite{blum1988non}.
Lastly, the remaining proofs, based on digital signature derivation, also show limited network use.

\subsection{Anonymization}
\label{exp:anonymization}
We have shown the feasibility of using different credential implementations with vastly different cryptographic primitives and now offset this to the anonymization component in \IPVAT{}.
We now measure latency as the time it takes to establish a communication channel between real users of \Tribler{}~\cite{pouwelse2008tribler}, our peer-to-peer file-sharing application that uses \IPVAT{}.
We present the creation time for channels of the three lengths default to \Tribler{}: using 1, 2 and 3 intermediary peers.
The more intermediaries a channel contains, the harder it will be for an adversary to decrypt or block any particular channel.

We have visualized the time it takes to create a channel for different intermediary counts in \autoref{fig:perfanon}, which shows that the parameterization of the anonymization component is very important for the total latency of the SSI-overlay interactions in \IPVAT{}.
Using 1-intermediary anonymization, the anonymization latency is hardly significant in comparison to the latency of the credential implementations (\autoref{fig:perflatency}).
When using two intermediaries, the anonymization latency is comparable to that of the credential implementations.
Finally, by using three intermediaries for anonymization, the latency is dominant compared to that of the credential implementations.
This result shows that the choice of the credential proof implementation is indeed insignificant for the latency of the credential data flows when compared to the anonymization component, using the standard of three intermediaries for covert channel construction.

\begin{figure}
\centering
\includegraphics[width=.7\columnwidth]{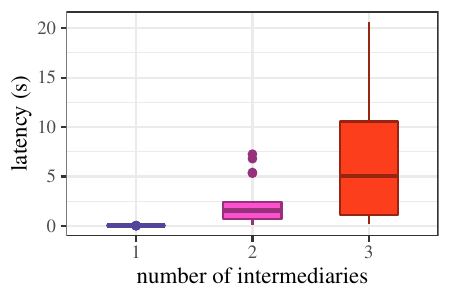} 
\caption{Covert channel creation time in \IPVAT{}.}
\label{fig:perfanon}
\end{figure}

\section{Related Work}

A promising start has been made in the field to try and make sense of all the available information disclosure technology, its mapping to the W3C DID standard~\cite{reed2020dids}, and its interaction with blockchains~\cite{muhle2018survey, kondova2020self} and distributed storage like IPFS~\cite{faisca2016decentralized}.
A critical view of the system components of Self-Sovereign Identity, however, is missing in academia and is only briefly discussed in industry whitepapers, the most notable analysis coming from SelfKey~\cite{selfkey2017selfkey}.
Many SSI solutions claim they are anonymous, but this is erroneous as they only achieve pseudonymity and do not address device fingerprinting~\cite{stopar2019emmy}.
We postulate that anonymity and identity disclosure protocols are orthogonal but can be combined to achieve pseudonymity with selective disclosure of information.

\textbf{Early work using central servers.} Early work focuses on supplying users with the ability to tie their claims to a token-based identity, as exemplified by Microsoft Passport~\cite{cameron2007design}, OAuth~\cite{chen2014oauth}, FIDO~\cite{lindemann2013evolution} and OpenID~\cite{recordon2006openid}.
At the time, the use-case of identification of users was to reidentify with the same central server---but these works laid the groundwork for cryptography, the terminology (e.g., ``claims'' and ``relying parties''), and the principles of Self-Sovereign Identity.

\textbf{Unlinkable signature-based disclosure schemes.}
Many mainstream SSI solutions are based on signature derivation (most notably Idemix's CL signatures~\cite{camenisch2002signature} and JSON Web Tokens~\cite{ethelbert2017json}), possibly storing revocations on blockchains~\cite{abraham2019privacy}.
These signatures bind the key of the identity holder to the disclosed data and usually allow derivation of a new signature that is also valid for the same data (which is claimed to make this data \emph{unlinkable}, though fingerprinting makes users completely linkable as we have previously discussed).
Examples of these systems are IRMA~\cite{alpar2017irma}, Jolocom and uPort~\cite{gruner2019integration}, ClaimChain~\cite{kulynych2018claimchain}, HyperLedger Indy and Sovrin~\cite{saraf2018blockchain}.

\textbf{ZKP-based identities.}
Identification through Zero-Knowledge protocols has been proposed decades ago~\cite{desmedt1988abuses} and these protocols have been rediscovered as a key component of Self-Sovereign Identities~\cite{baars2016towards}.
However, Zero-Knowledge Proofs over data and their security implications for SSI systems remain scarcely documented and rarely implemented by academia~\cite{borse2019anonymity}, though this type of system is widely proposed by industry, with \url{https://github.com/peacekeeper/blockchain-identity} listing $134$ Self-Sovereign Identity solutions based on blockchain.

\section{Summary}

Research on Self-Sovereign Identities is narrowly focused on cryptographic data disclosure protocols.
However, our analysis shows that these protocols are not the only critical consideration for Self-Sovereign Identity solutions.
Such an isolated approach ignores performance and security concerns at the networking layer of SSI systems.
We have presented our implementation of a system that addresses these concerns.
Furthermore, we have shown the feasibility of a passport grade Self-Sovereign Identity through our design and implementation of TCID.
Our Self-Sovereign Identity solution of TCID disallows network traffic correlation, while still enabling passport-grade interactions with acceptable latency.

\section*{Availability}
All code of \IPVAT{} is available on GitHub at \IPVATGITHUB{} under the GNU LGPL 3.0 license.

\bibliographystyle{IEEEtranS}
\bibliography{bibliography}

\end{NoHyper}
\end{document}